\documentclass[10pt,aps,prb,twocolumn,amsmath]{revtex4-1}
\usepackage{graphicx}
\usepackage{dcolumn}
\usepackage{bm}
\usepackage{amsmath}
\usepackage{color}
\begin{document}

\title{First-principles study of $\langle c+a \rangle$ dislocations in Mg}
\author{Anil Kumar$^{1}$, Benjamin Morrow$^{2}$, Rodney J. McCabe$^{2}$ and Irene J. Beyerlein$^{1}$}
\affiliation{$^{1}$ Theoretical Division, Los Alamos National Laboratory, Los Alamos, NM 87545 \\
                $^{2}$ Materials Science and Technology Division, Los Alamos National Laboratory, Los Alamos, NM 87545}           
\date{\today}

\begin{abstract}

We use first-principles density functional theory to study the generalized stacking fault energy surfaces for pyramidal-I and pyramidal-II slip systems in Mg.  We demonstrate that the additional relaxation of atomic motions normal to the slip direction allows for the appropriate local minimum in the generalized stacking fault energy (GSFE) curve to be found. The fault energy calculations suggest that formation of pyramidal-I dislocations would be slightly more energetically favorable than that for pyramidal-II dislocations. The calculated pyramidal-II GSFE curves also indicate that the full pyramidal II dislocations would dissociate into the Stohr and Poirier (SP) configuration, consisting of two $\frac{1}{2}\langle c+a \rangle$ partials, $\frac{1}{6}[11{\bar2}3] + \frac{1}{6}[11{\bar2}3]$ , but the pyramidal-I GSFE curves, while also possessing a local minimum, would not dissociate into the same SP configuration. We report observation of these partials here emanating from a $\{10{\bar1}2 \}$ twin boundary. Using MD simulations with MEAM potential for Mg, we find that the full pyramidal-II $\langle c+a \rangle $ dislocation splits into two equal value partials $\frac{1}{6}[11{\bar2}3] + \frac{1}{6}[11{\bar2}3]$ separated by ~22.6 $\AA$.  We reveal that the full pyramidal-I $\langle c+a \rangle$ dislocation dissociates also into two equal value partials but onto alternating $(30{\bar3}4)$ and $(30{\bar3}2)$ planes with $\frac{1}{6} [20{\bar2}3]$ and $\frac{1}{6} [02{\bar2}3]$ Burgers vectors separated by a 30.4 $\AA$ wide stacking fault.  When a stress is applied, edge and mixed dislocations of the extended pyramidal-II dislocation can move on their glide plane; however, pyramidal-I dislocations of similar character cannot.  

\end{abstract}

\maketitle

\section{Introduction} 

Lightweight structural materials have many applications in the aerospace and automotive industries as they can drastically decrease energy consumption and improve fuel economy. Magnesium (Mg) and its alloys are some of the lightest structural materials among metallic alloys, but many compositions exhibit very low ductility and highly anisotropic mechanical behavior.  Mg has a hexagonal close packed (HCP) structure that lacks the ductility and formability of cubic materials, due to the scarcity of easy crystallographic slip systems to accommodate an arbitrary plastic deformation mode\cite{Lu-1,Lu-2}.  With the idea that a better fundamental understanding can help overcome these deficiencies, the mechanisms of slip operating within these lightweight material candidates have been actively investigated both experimentally and computationally for the past few decades \cite{Pollock, Cole, Bauer, Williams, Kumar-acta, Kumar-apl, HCWu}.   

The easiest slip mode in Mg is basal slip; however, this mode only provides two independent slip systems \cite{Wonsiewicz, RobertCS}.  A key to achieving formability in Mg is to increase the relative contributions of the non-basal slip systems to crystal deformation.  It is believed that deformation along $\langle c \rangle$ is accomplished by glide of $\langle c+a \rangle $ dislocations on the pyramidal slip planes. There are two types of pyramidal $\langle c+a \rangle$  slip systems that have been observed or proposed to accommodate $\langle c \rangle$ axis deformation:  pyramidal type I $(10{\bar1}1) \langle 11{\bar2}3\rangle $ and type II $(11{\bar2}2) \langle 11{\bar2}3\rangle $.  These glide on different planes exhibit different atomic densities and hence are not expected to have the same activation barriers, dislocation cores, and response to solute atoms.  To date, which type of slip system prevails in Mg and its alloys is not very well understood.  

Over the years, experimental, analytical and numerical methods have been used to infer the dominant pyramidal slip mode in Mg. Experimental studies using slip traces on deformed single crystals or crystals embedded in grains have provided evidence for both types of pyramidal slips \cite {Obara, Agnew1, Byer, Lilleodden, Kelvin, Sandlobes}. Similarly, post-mortem TEM analyses on deformed Mg samples have reported $\langle c\rangle $ and  $\langle c+a \rangle $ dislocations on both pyramidal planes\cite{Kelvin, Obara, Syed, Geng}.  The dislocations found after deformation were both of the glissile and sessile type.  Based on experimental observation, Stohr and Poirier \cite{Stohr-1972} proposed that full pyramidal-II dislocations can lower their energy by splitting into two equal length 1/2 $\langle c+a \rangle $ partials that spread apart while remaining on the pyramidal-II plane.  A pyramidal-II dislocation with such a core configuration would likely have been glissile.  Recently, Yu et al. studied the deformation of Mg single crystals of nano-scale dimension and witnessed the glide of a pyramidal type II dislocation in-situ \cite{Yu-Qian-1}. 

Atomic-scale MD calculations have not always produced results consistent with experimental observation.  Whether the lower energy SP configuration occurs has proven to depend on the interatomic potential and temperature.  Using Lennard-Jones potential, Ando et al. \cite{Ando-S} found that full dislocation dissociated into the SP glissile configuration at 0K and room temperature but not at 30K, wherein it dissociated into a sessile configuration of two partials connected by a basal stacking fault.  The latter was proven sessile under a shear applied to the glide plane and direction.  Using a larger simulation cell and two other potentials, the embedded atom method (EAM) and Finnis-Sinclair (FS) potentials, Morris et al.\cite{Morris} found that the full pyramidal-II dislocation dissociated into another similarly sessile configuration.  More recently, with a modified embedded atom method (MEAM) potential, Wu et al. \cite{Curtin1} studied the stable configuration of two $  \frac{1}{2}\langle c+a \rangle $ partials on pyramidal I and II planes.  Later, Wu and Curtin \cite{CurtinNature} using finite-temperature MD showed that via a thermally activated process ($>= 500$ K), the SP configuration for both pyramidal type I and II dislocations could transform into an even lower energy state, which is sessile.  The reconstructed core of the pyramidal-I dislocation achieved the lower energy state and hence was considered the less likely of the two pyramidal types to mediate plasticity.  As an alternative approach, dynamic loading MD simulations have been carried out to investigate which of the two types of pyramidal slip dislocations would preferably nucleate.  Among these studies, either both types formed or just the formation of pyramidal type I dislocations was observed \cite{DH-Kim, BLi, Alwady, Guo}.  


\begin{figure}
\includegraphics[scale=0.75]{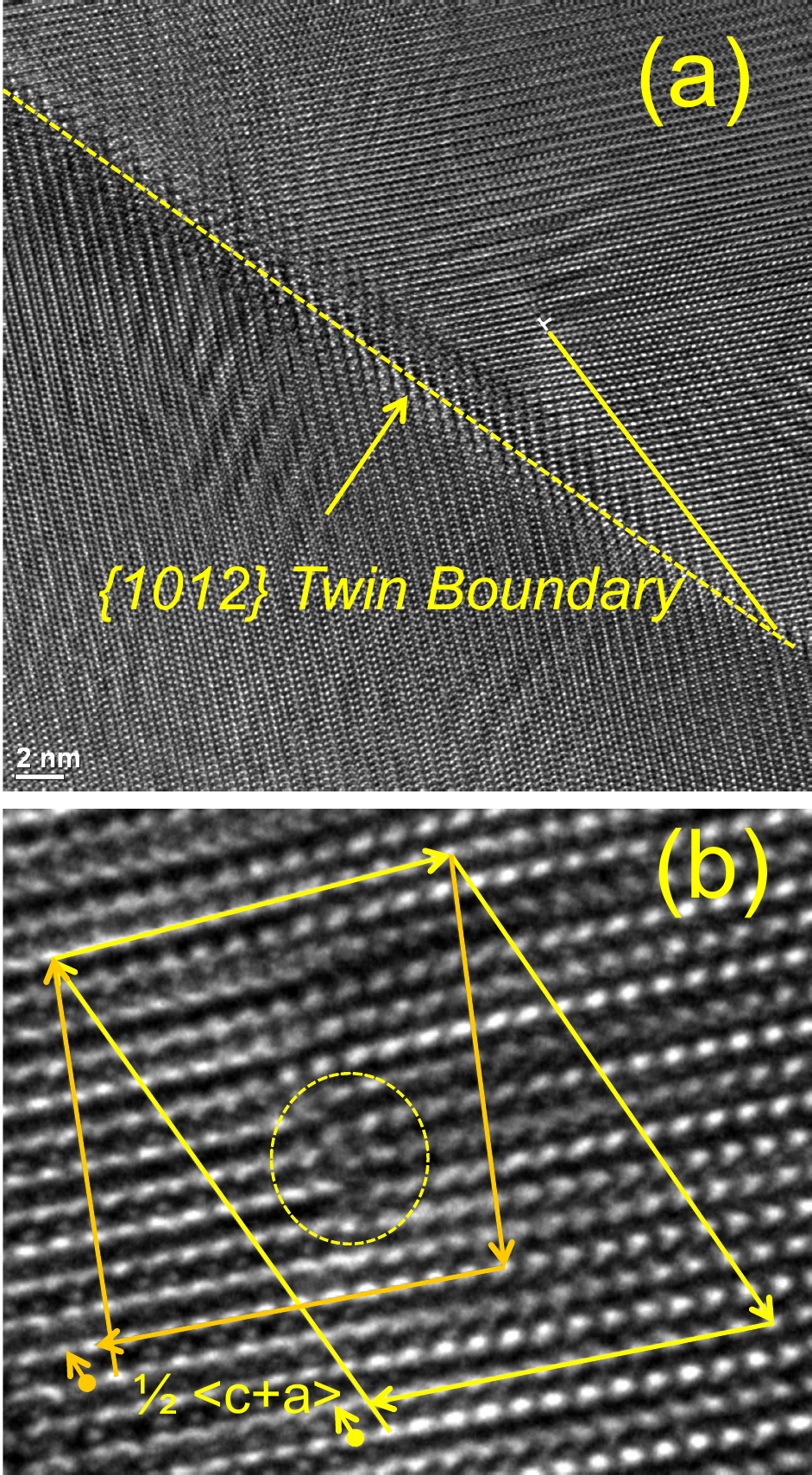}
\caption{High resolution transmission microscopy on a $\langle 11{\bar 2}0 \rangle$ zone axis showing the presence of a $\frac{1}{2} \langle c+a \rangle$ partial dislocation.  (a) position of the dislocation and trace of the pyramidal plane extending to the (10${\bar1}$2) twin boundary.  (b) Burgers circuits drawn around the dislocation showing the Burgers vector perpendicular to the viewing direction to be $\frac{1}{2} \langle c+a \rangle$.  The stacking fault is not evident in the HRTEM from this zone axis, which is consistent with previous DFT modeling and our DFT modeling. }
\label{expt-result}
\end{figure}

\begin{figure*}
\includegraphics[scale=0.60]{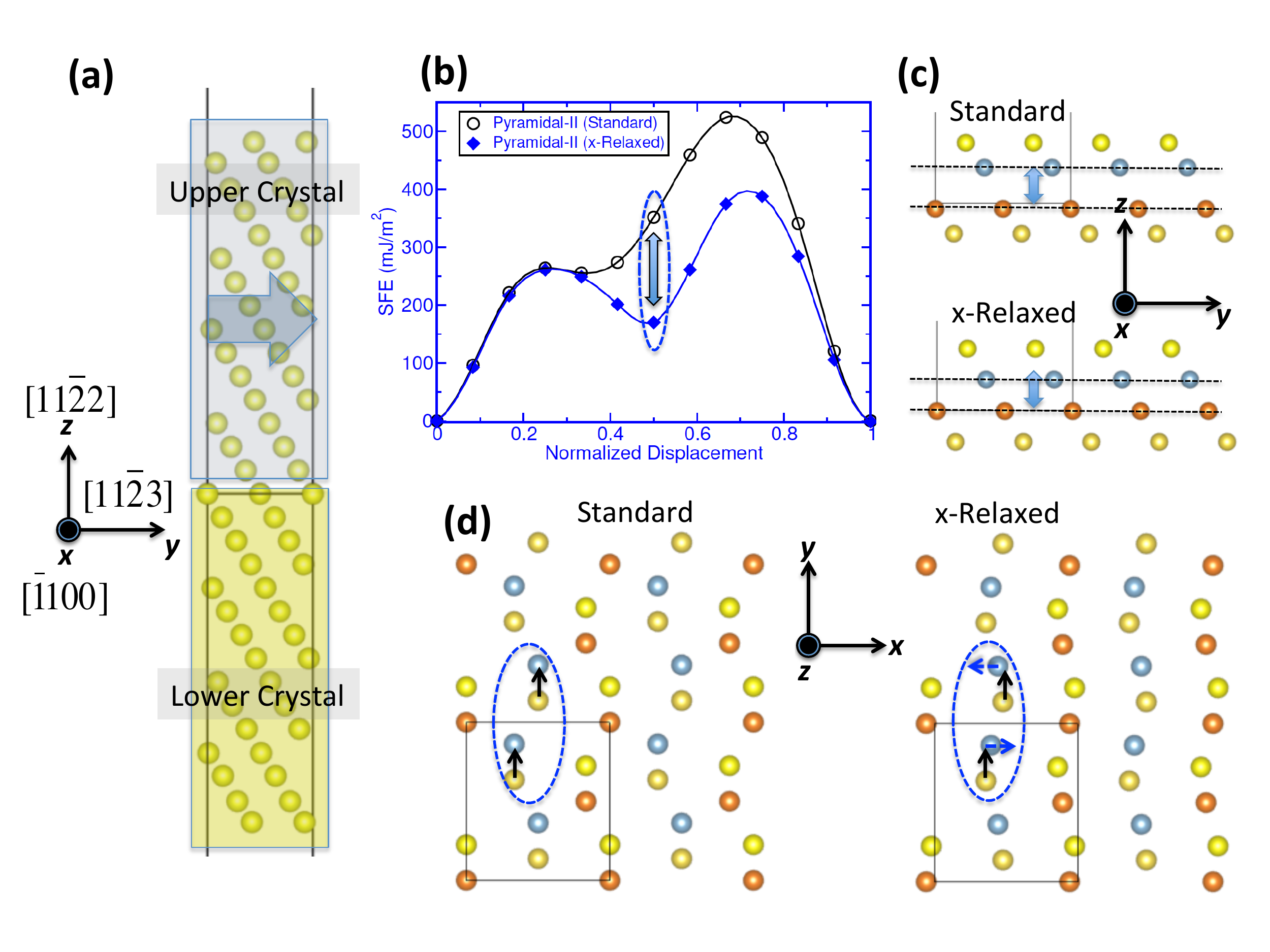}
\caption{The GSFE curve for pyramidal-II slip system in Mg obtained from the density functional theory calculations. (a) shows the crystallographic orientation of the periodic supercell. (b) shows the calculated GSFE curve using the standard and x-Relaxed approaches as discussed in the text. The panel (c) compares the change in interlayer spacing along the z direction for the standard and x-Relaxed methods when the normalized displacement is x = 0.5. Panel (d) shows the local shuffling of atoms in the x-direction near the slip planes for the two approaches.}
\label{slip-py2}
\end{figure*}

\begin{figure*}
\includegraphics[scale=0.60]{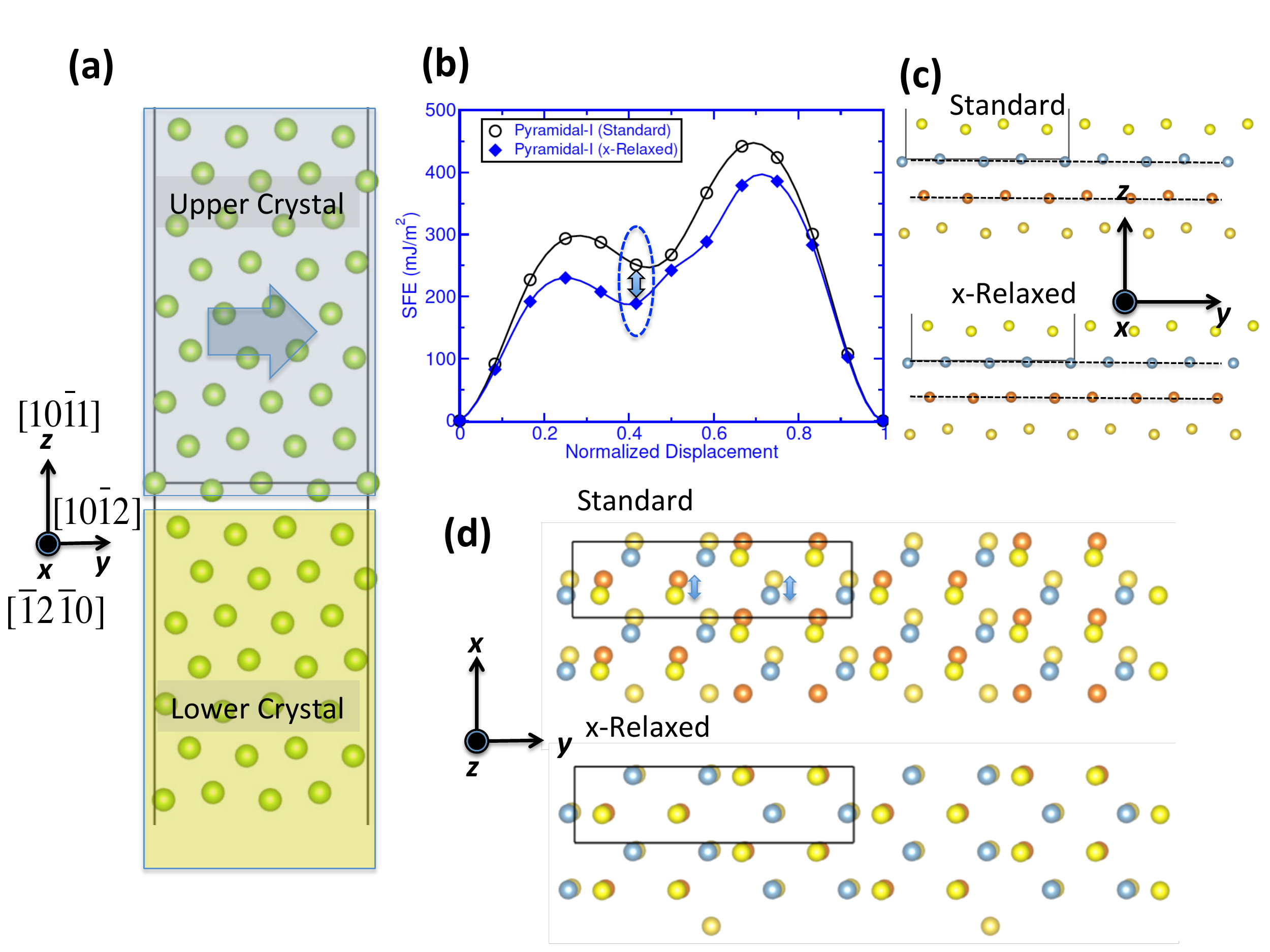}
\caption{The GSFE curve for the pyramidal-I slip system in Mg obtained from density functional theory calculations. Panel (a) shows the crystallographic orientation of the periodic supercell. To calculated the GSFE curve, we move the upper crystal along $\frac{1}{3}[11{\bar2}3]$ direction. Panel (b) shows the calculated GSFE curve using the two approaches. Panel (c) compares the change in interlayer spacing along the z direction for the standard and x-Relaxed methods and for the normalized displacement of x = 0.41. Panel (d) shows the local shuffling of atoms in the x-direction near the slip planes for the two approaches.}
\label{slip-py1}
\end{figure*}


Ab initio density functional theory (DFT) has also been applied to this problem and for a pyramidal-II dislocation, the SP dissociation was observed \cite{Curtin2}.  Most often, however, DFT is used to calculate the generalized stacking fault energy (GSFE) curve \cite{Alwady, Curtin1}.  The GSFE curve is obtained by rigidly displacing one half of the crystal relative to the other half across a plane. From this energy profile, the lattice resistance to move the dislocation and the possible set of partial dislocations that can produce the same displacement as a full dislocation can be estimated. To be consistent with the SP dissociation seen in DFT and MD simulations, the corresponding GSFE curve on the $\{11{\bar2}2\}$ plane with displacements in the $\langle 11{\bar2}3 \rangle$ direction should possess a local minimum at $\frac{1}{2} \langle c+a \rangle$.  However no such DFT-calculated or MD-calculated GSFE curve to date exhibits this local minimum.  The local minimum in the GSFE curve was thought to either lie elsewhere on the $\gamma$ surface or not exist \cite{ Alwady, Nogaret}.  Atomic-scale simulations are well suited for providing valuable insight into pyramidal $\langle c+a \rangle$ dislocation properties yet the inconsistencies in core structure and dislocation dynamics reported among so many atomic-scale studies clearly need to be addressed. 

In this work, we study the core configuration and glide behavior of pyramidal type I and type II dislocations under stress.  Direct experimental evidence is provided for a single  $ \frac{1}{2} \langle c+a \rangle$ dislocation on the pyramidal type II $\{11{\bar2}2  \}$ plane. Using DFT we calculate the GSFE curve for both type I and II pyramidal slip systems.  In calculation, we show that proper boundary conditions are needed to allow atoms in the pyramidal planes to shuffle to reach a local minimum.  It is then demonstrated that a minimum exists exactly at $ \frac{1}{2} \langle c+a \rangle$  along the $\langle 11{\bar2}3 \rangle$ slip direction in the GSFE curve for the pyramidal type II dislocation, consistent with the SP reaction and experimental evidence, and supporting the SP dissociations calculated by some MD studies. With these relaxed conditions, we show that GSFE curves from three Mg interatomic potentials also exhibit a local minimum at the amount of same displacement.  We also simulate the motion of edge and mixed pyramidal type I and II dislocations using MD simulation.  The calculations reveal that stress alters the core structure and the motion of each type in the glide plane is anisotropic depending on the sense of shearing direction.  For extended dislocations of edge or mixed character, we show that both partials of the extended pyramidal type II dislocation can glide on the glide plane, whereas only one partial of the extended pyramidal type I can glide.  

\section{Methods}

\noindent {\bf Experiment:} In order to generate pyramidal dislocations, highly textured pure polycrystalline Mg was lightly deformed such that the c-axes of most crystals were extended.  Transmission electron microscopy (TEM) foils were prepared by sectioning and chemically thinning to 150 $\mu$ m.  Three mm discs were punched and electropolished to electron transparency using $2\%$ nitric acid in water and minimal voltage.  HRTEM was performed in an image-corrected FEI Titan with an accelerating voltage of 300 kV.  Figure~\ref{expt-result} shows an HR-TEM image of a partial dislocation on the pyramidal $\{11{\bar2}2 \}$ plane that emanates from a $\{10{\bar1}2 \}$ twin boundary. A Burgers circuit analysis shows that the Burgers vector of this partial dislocation is $ \frac{1}{2} \langle c+a \rangle$.  This analysis provides evidence of a $\frac{1}{2} \langle c+a \rangle$  partial dislocation on the $\{11{\bar2}2\}$ plane.

\noindent {\bf Density Functional Theory:} We carried out density functional theory (DFT) based calculations using generalized gradient approximation (GGA) for the exchange correlation functional with the Perdew-Becke-Erzenhof (PBE) parametrization\cite{perdew} as implemented in the VASP code \cite{vasp1, vasp2}. The interaction between valence electrons and ionic cores is treated using PAW pseudopotentials \cite{paw1,paw2}. The number of valence electrons in the Mg pseudopotential is two (3s$^{2}$). In our DFT calculations, we used a plane wave energy cutoff of 350 eV, and optimized the atomic structure until the force on each atom is smaller than 0.01 eV/$\AA$. We used a $19\times 19\times 11$ $\Gamma$-centered Monkhorst Pack \cite{mh-pack} k-point to integrate the Brillouin Zone of the primitive unit cell to calculate the lattice constants. The resulting values are $a=3.19 \AA$ and $c=5.187 \AA$, which are in very good agreement with measurements. These values are then used to construct the supercell to study the generalized stacking fault energy curve for the pyramidal-I and pyramidal-II slip systems. 

\noindent{\bf  Molecular Dynamics Simulations:} We performed the Molecular Statics (MS) and Molecular Dynamics (MD) simulations using the Large-scale Atomic/Molecular Massively Parallel Simulator (LAMMPS) \cite{lammps}. In our MD simulations, we used a recently developed modified embedded-atom method (MEAM) potential by Wu et al. \cite{Curtin1} for the Mg. To obtain the relaxed structure in the MS simulation we optimize the total energy of the system using the conjugate gradient method. In the MD simulations at finite temperature, we used a constant volume ensemble with velocity verlet algorithm to integrate the equations of motion to sample equilibrium configurations. 

\section{Results and Discussion}

\subsection{Generalized Stacking Fault Energy}

To calculate the generalized stacking fault energy curve using density functional theory, we use periodic boundary conditions in the x and y directions. For the pyramidal-II slip system, the periodic simulation cell and coordinate system are shown in Figure~\ref{slip-py2}(a).  Our periodic model contains 60 atoms and the dimensions are 5.53 $\AA$ along x, 6.088 $\AA$ along y and 51.49 $\AA$ along the z, which spans 30 atomic layers. 

Following the standard approach used for GSFE calculations, we shift the upper half of the crystal with respect to the lower half of the crystal along the $\langle 11 {\bar 2} 3 \rangle $ direction. At each displacement, we minimize the energy of the system by fixing all the positions of atoms in both the upper and lower crystals in the x and y directions and allow the positions in the z direction to relax. The calculated GSFE curve for the pyramidal-II $(11 {\bar 2} 2) \langle 11 {\bar 2} 3 \rangle$ slip using this standard approach is shown in  Figure~\ref{slip-py2}(b). The displacement on the x-axis in the Figure~\ref{slip-py2}(b) is normalized by the periodic length along the slip direction, which corresponds to the full $\langle c+a \rangle$ dislocation. The standard GSFE for the pyramidal-II slip shows a very shallow minimum at 0.33 normalized displacement. This calculation does not suggest a dissociation of a full dislocation into two 1/2 $\langle c+a \rangle$ partials. It suggests that the Burgers vectors for a pyramidal-II dislocation would be ($\frac{1}{9} [11{\bar 2}3 ]$  and $\frac{2}{9} [11{\bar 2}3 ]$), which disagrees with prior experimental observations, our present experimental observation, and some atomic scale simulations of dislocation dissociations. 

The standard approach does not permit the appropriate local minimum to be found for achieving the $\langle c+a \rangle$ dislocation partials on the type II plane \cite{Chaari}.  To rectify this problem, we used another approach to calculate the GSFE curve, where we only fixed the atomic positions in the y direction in the both upper and lower crystals and allowed all the atoms to relax in the x and z directions. The calculated GSFE curve using this second approach, denoted as the as x-Relaxed approach, is shown in Figure~\ref{slip-py2}(b). Interestingly, when we allow atoms to relax in the perpendicular direction (i.e., along x) while displacing the upper crystal in the y direction, we find a local energy minimum at 0.5 of the normalized displacement. This value of normalized displacement corresponds to $\frac{1}{2}\langle c+a \rangle$ partial dislocation with Burgers vector {$\frac{1}{6} [11{\bar 2}3 ]$}, in agreement with experimental evidence.


\begin{figure*}
\includegraphics[scale=0.70]{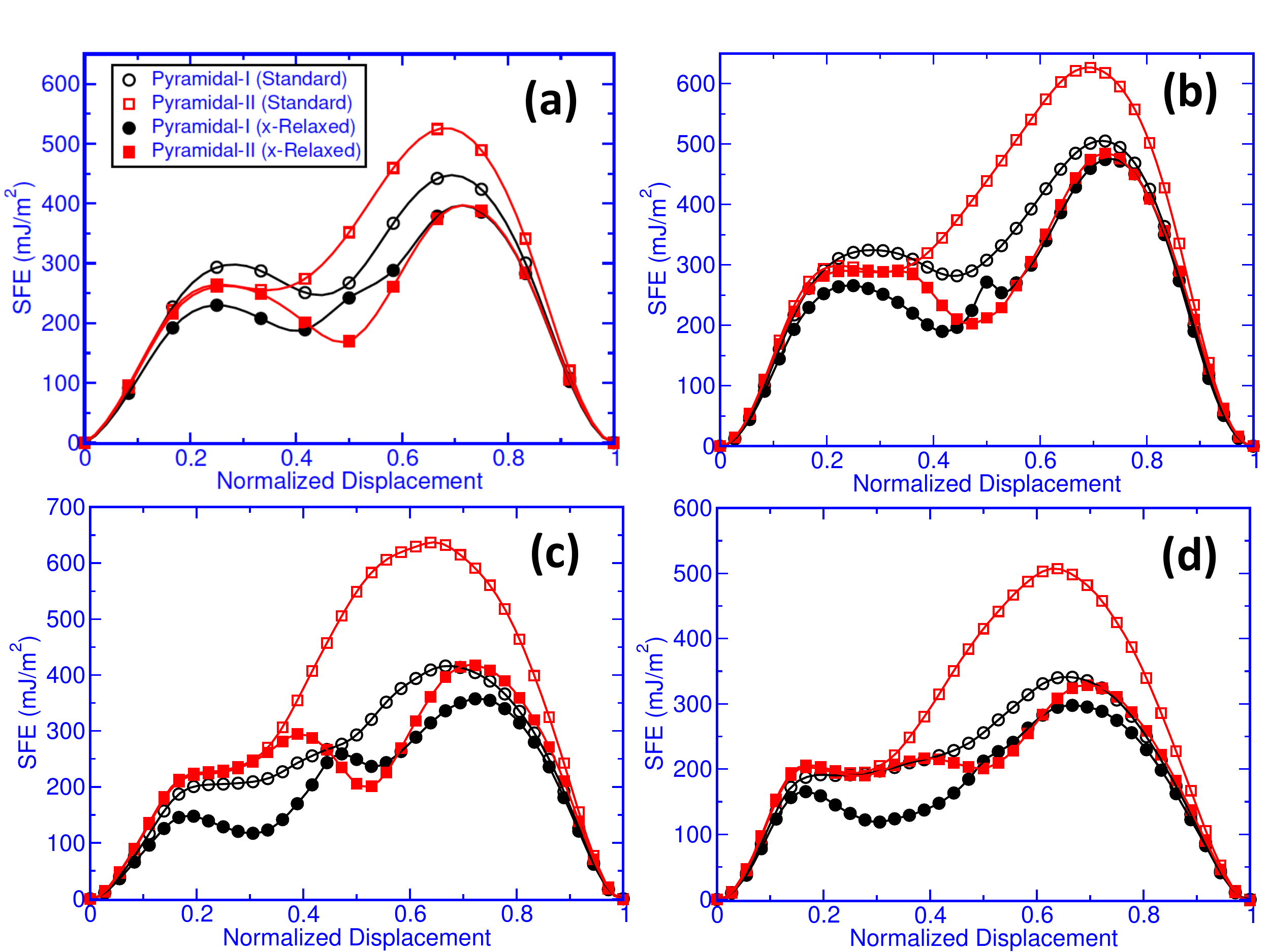}
\caption{Comparison of the calculated GSFE curves for the pyramidal-I and pyramidal-II slip systems using the standard and x-Relaxed approaches from DFT and MD. Panel presents results from (a) DFT, (b) the MEAM potential, (c) the Sun\cite{Sun-EAM} EAM potential and (d) the Liu \cite{Liu-EAM} EAM potential.}
\label{slip-py1-py2}
\end{figure*}

\begin{figure}
\includegraphics[scale=0.35]{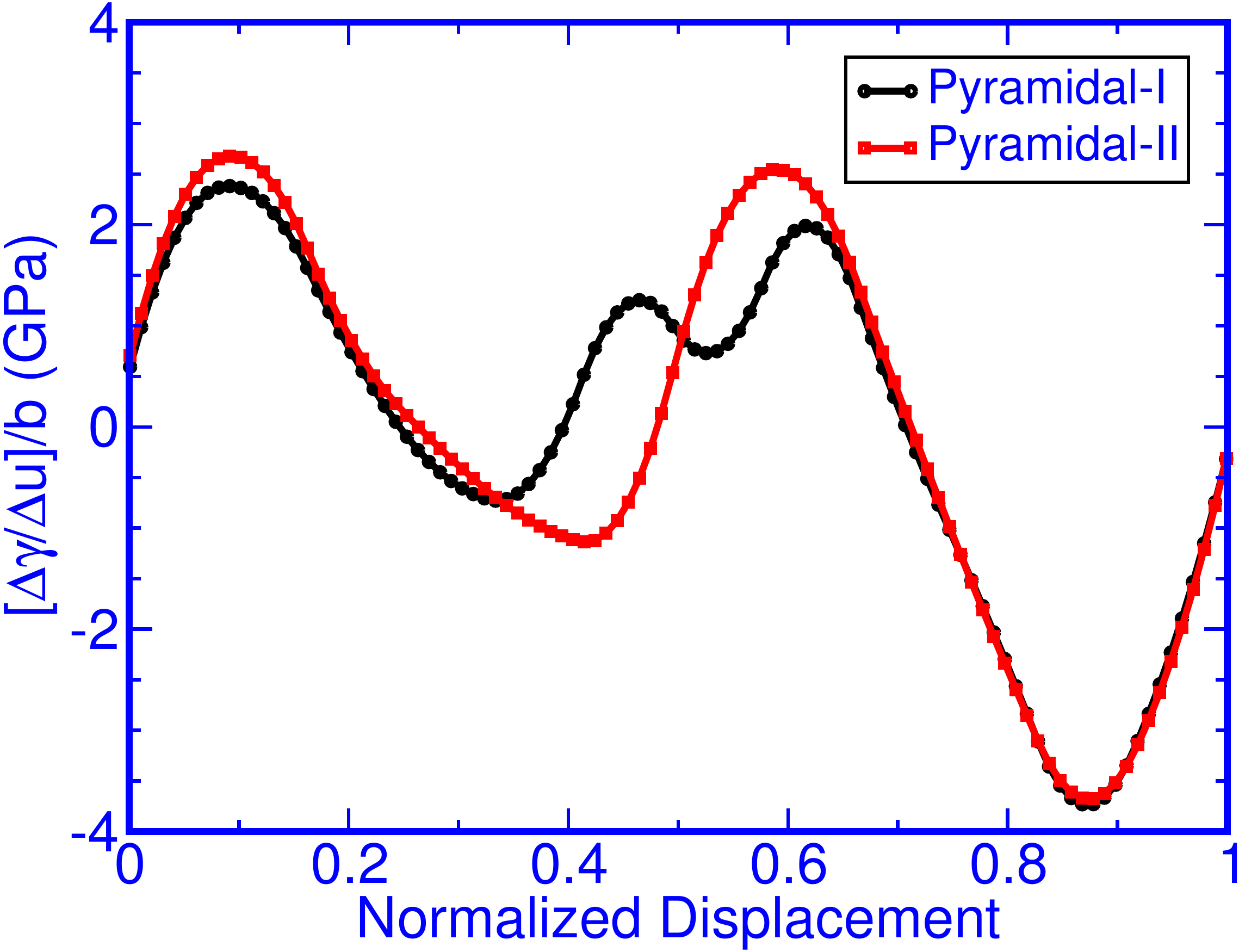}
\caption{The derivative $\frac{1}{b}\frac{\partial \gamma} {\partial u}$ corresponding to the x-Relaxed GSFE curves for the pyramidal-I and pyramidal-II slip systems.}
\label{Joos}
\end{figure}

\begin{figure*}
\includegraphics[scale=0.70]{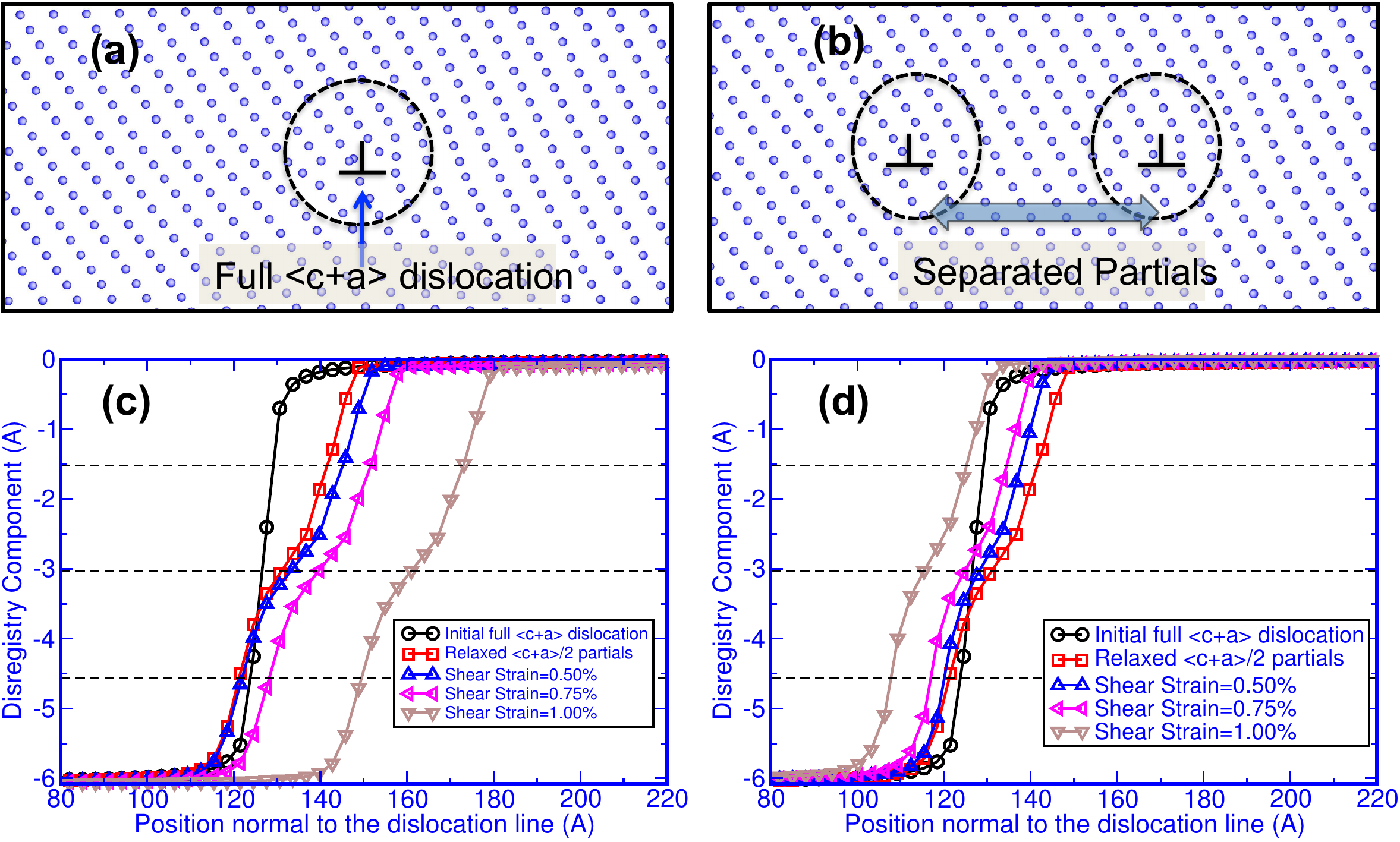}
\caption{Dissociation of a $\langle c+a \rangle$ edge dislocation into two $\frac{1}{2} \langle c+a \rangle $  partials on the pyramidal-II plane. (a) shows the initial full $\langle c+a \rangle$ edge dislocation on the $(11 {\bar 2} 2)$ slip plane, which after relaxation splits into two  $\frac{1}{2}\langle c+a \rangle$ partials as shown in (b). The (c) and (d) are disregistry plots that reveal the effect of the positive and negative shear strain parallel to the $(11 {\bar 2} 2)$ plane on the further splitting and motion of these two partials.}
\label{splitting-edge-py2}
\end{figure*}

\begin{figure}
\includegraphics[scale=0.6]{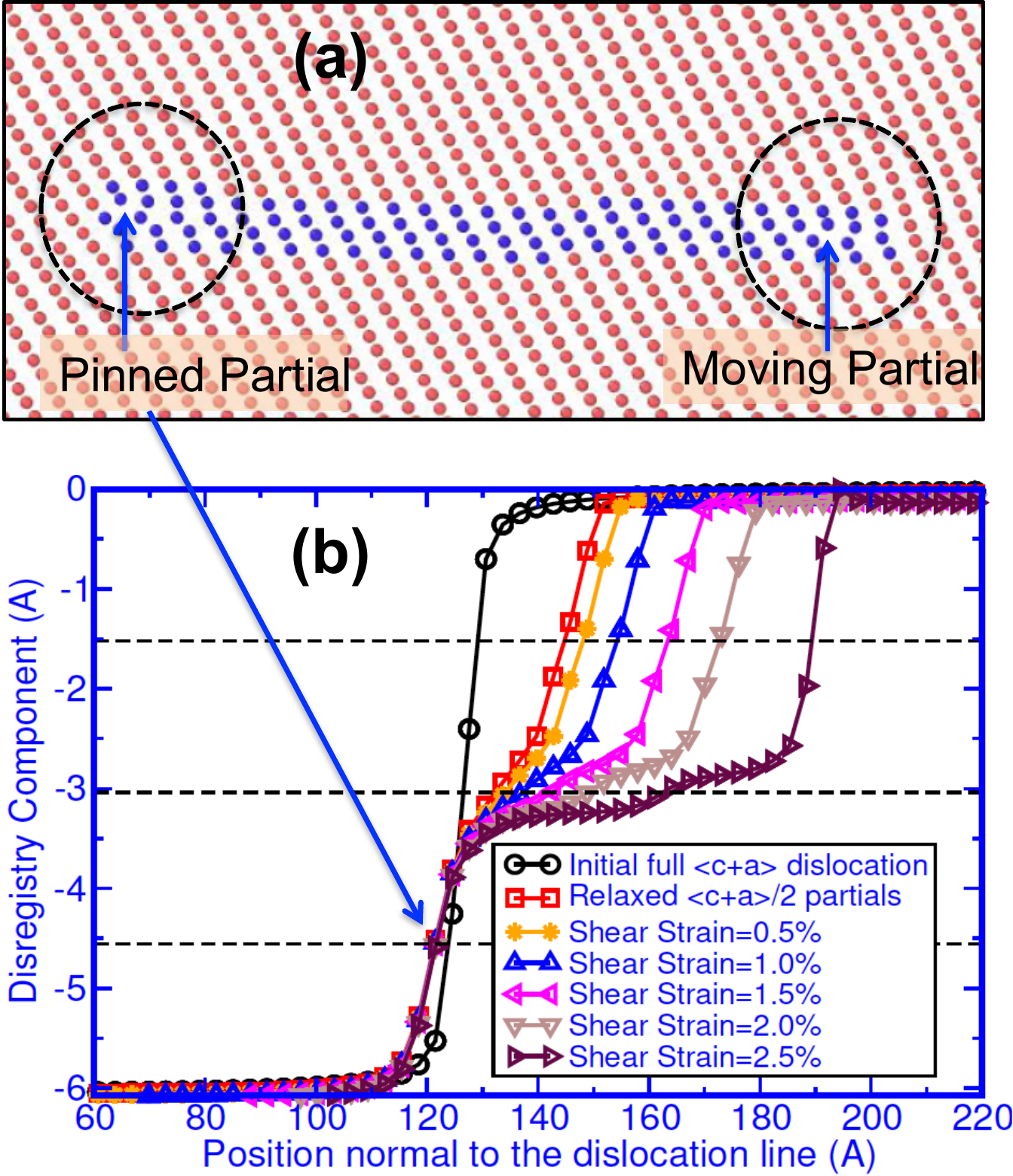}
\caption{Further splitting of two $\frac{1}{2}\langle c+a \rangle$  partials under the applied shear strain when one of them is pinned: (a) shows the structure when the partial on the right has moved far away on the pyramidal-II plane with applied shear strain when the other partial on the left is pinned. (b) disregistry plots to show the separation of two partials under the applied shear strain}
\label{explain-expt-result}
\end{figure}

\begin{figure*}
\includegraphics[width=18cm,height=14cm,keepaspectratio]{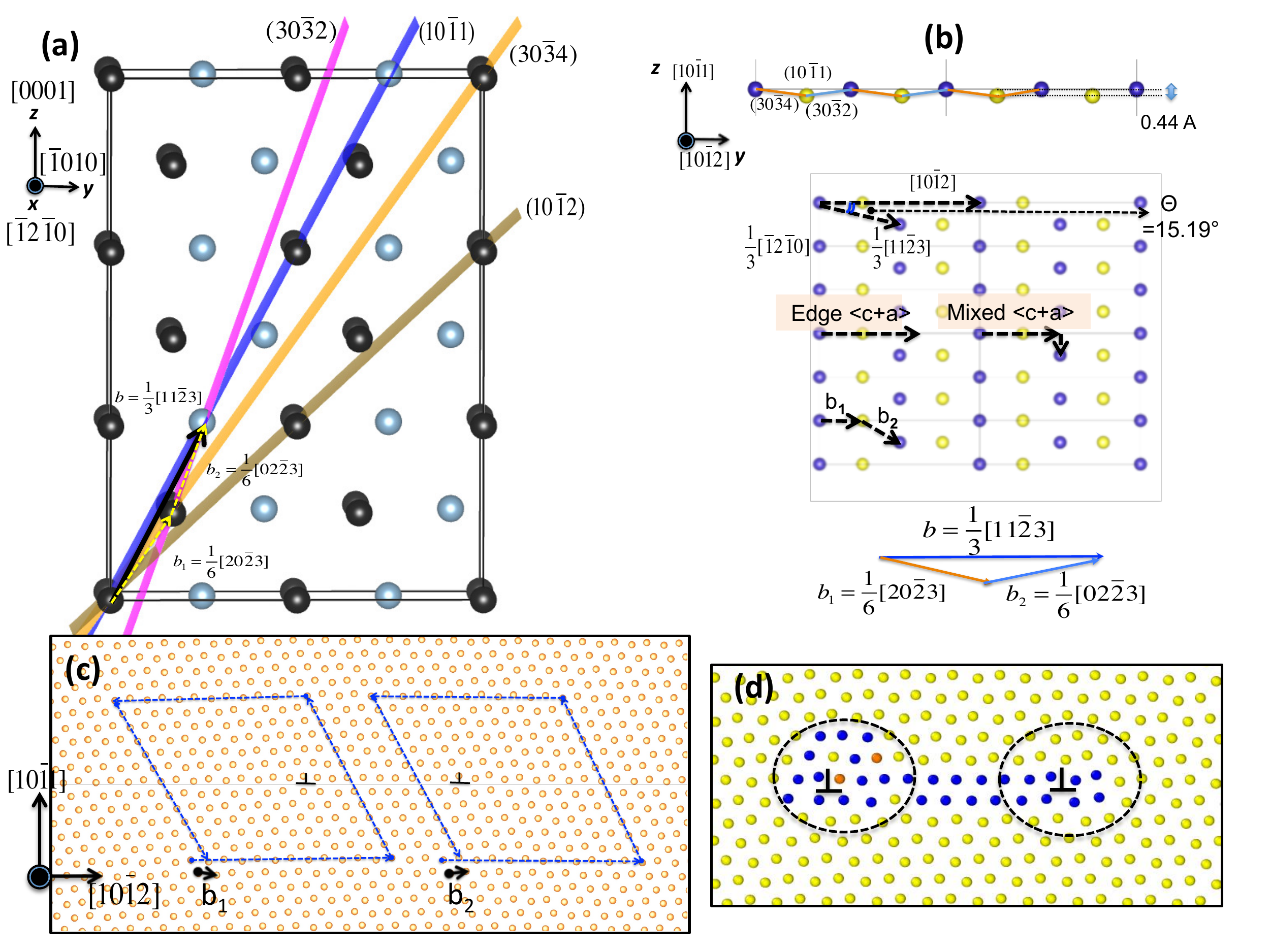}
\caption{(a) Orientation of different crystallographic planes in an HCP crystal. Atoms in the x=0 plane are shown in black and atoms in the x=$\frac{1}{2}$ plane are shown in silver. Panel (b) shows the side and top view of the ($10{\bar1}1$) plane. The ($10{\bar1}1$) is serrated (consists of two atomic layers shown as atoms in blue and yellow), and can be considered made of locally of the $(30{\bar3}4)$ and $(30{\bar3}2)$ planes as shown in (a). The $\langle c+a \rangle$ dislocation Burgers vector $\frac{1}{3} [11{\bar2}3]$ on the ($10{\bar1}1$) plane can split into two partials of Burgers vectors $\frac{1}{6} [20{\bar2}3]$ and $\frac{1}{6} [02{\bar2}3]$.  (c) shows the Burgers circuit for the partials obtained from the dissociation of the $\langle c+a \rangle$ dislocation on the ($10{\bar1}1$) plane in our MD simulations. (d) shows the partial core sturucture and the stacking fault region between the two partials. }
\label{splitting-py1}
\end{figure*}

\begin{figure}
\includegraphics[scale=0.3]{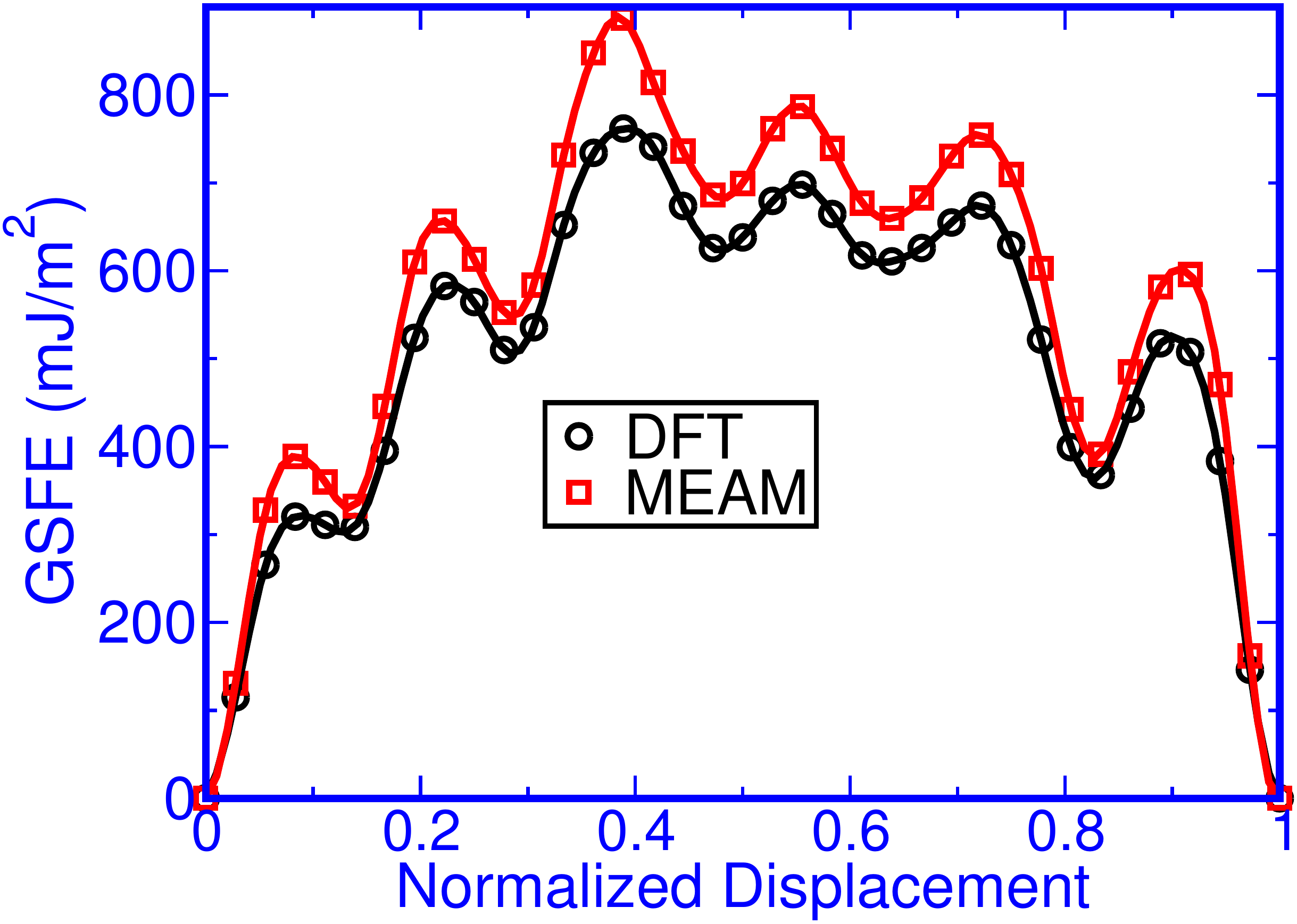}
\caption{Presence of a local minimum at $\frac{1}{6} [20{\bar2}3]$ on the GSFE curve for the $(30{\bar3}4) \langle 20{\bar2}3 \rangle $ obtained from both DFT and MD.}
\label{gsfe3034}
\end{figure}


To understand the mechanism for energy minimization in the x-Relaxed approach compared to the standard approach, we compared the relaxed structures obtained from the two approaches at 0.5 of normalized displacement.  Figure~\ref{slip-py2}(c) shows the interlayer spacing of the two atomic layers from the upper crystal and two atomic layers from the lower crystal of the relaxed structures calculated from the two approaches. In the standard approach, we find that interlayer spacing increases as we displace the upper crystal with respect to the lower crystal. The interlayer spacing at the interface of the upper and lower crystal for a displacement equal to 0.5 is 1.86 $\AA$ compared to the bulk interlayer spacing 1.36 $\AA$. The local shuffling of the atoms in the x direction in the x-Relaxed approach compared to the standard approach is shown in Figure~\ref{slip-py2}(d). We observe that the relaxation of atomic positions along the x direction in the x-Relaxed approach allows atoms to locally shuffle, altering their positions in order to minimize the interlayer spacing at the interface. Evidently, the shuffling mechanism for shearing on pyramidal planes is critical for minimizing the total system energy.

Similarly, we studied the GSFE for pyramidal-I slip using the standard and x-Relaxed approaches. The model for the pyramidal-I slip is shown in Figure~\ref{slip-py1}(a) and contains 64 atoms and the periodic dimensions are 3.19 $\AA$ along x, 11.75 $\AA$ along y and 56.49 $\AA$ along the z spanning 16 atomic layers.  Because of the double lattice structure, the $\{10{\bar1}1\}$ planes have two interplanar spacings, 0.41 $\AA$ within a corrugated plane and 2.02 $\AA$ between two corrugated planes.  Here the displacements are applied on the atomically dilute plane that lies inbetween two corrugated $\{10{\bar1}1\}$ planes.  We find that shearing on denser plane in the direction of slip requires much more energy than the dilute one. The calculated GSF energies for the pyramidal-I slip (i.e., $(10 {\bar 1} 1) \langle 11 {\bar 2} 3 \rangle$) using the two approaches is show in Figure Figure~\ref{slip-py1}(b). The relaxation of atomic positions in the x directions minimizes the total energy of the system compared to the standard approach and shifts the local minimum to a normalized displacement of 0.4. Repeating the analysis of the atomic positions near the interface for the pyramidal type I plane shows that the main mechanism for lowering the energy of the system for the x-Relaxed approach is local shuffling of atoms in the x direction, as shown in Figure~\ref{slip-py1}(c) and (d). 

The local minimum in the x-Relaxed GSFE curve implies that this dislocation can find a lower energy state by splitting into smaller partial dislocations.  However, because the local minimum does not lie at 0.5, the production of $\frac{1}{2}\langle c+a \rangle$ partial dislocations on the $(10 {\bar 1} 1)$ plane is not likely. Most MD studies and topological analyses have proposed pyramidal-I slip dissociations into partial dislocations with unequal Burgers vectors \cite{ Alwady, Bacon, BLi, DH-Kim, Jones} 
an outcome that would be consistent with our x-Relaxed GSFE calculation. In many of these works, the Burgers vector of the leading partial was not aligned with the $(10 {\bar 1} 1) \langle 11 {\bar 2} 3 \rangle$ direction.  TEM provided evidence of a $\frac{1}{6} [20{\bar2}3] + \frac{1}{6} [02{\bar2}3]$ split although the pyramidal plane type was not specified \cite{ Sandlobes}. Prior MD simulations have shown that this extended dislocation can form in stressed crystals \cite{Guo, Q-Zu}. Alternatively, MD simulations by Li and Ma \cite{BLi} suggests that the leading partial corresponds to a twinning dislocation with its Burgers vector aligned along [10${\bar1}$2]. Hence, it may be the case that the local minimum lies outside of the GSFE curve for $(10 {\bar 1} 1) \langle 11 {\bar 2} 3 \rangle$ displacements. 

Some useful implications can be gleaned from comparing the x-Relaxed GSFE curves for pyramidal type I and II slip systems.  Both curves are asymmetric, suggesting that if the dislocations are glissile, then the response to glide depends on the sense of shear.  Both curves also exhibit a local minimum along the $\langle c+a \rangle$ direction, suggesting that a dissociated state for the full dislocation would be lower in energy.  Pyramidal I has a lower peak value than pyramidal II, indicating that it would be easier to form or glide. To relate the fault energy to a critical shear stress, we calculate the derivative ($\frac{1}{b} \frac{\partial \gamma} {\partial u} $) corresponding to the x-relaxed GSFE curves\cite{Joos-Duesbery}.  Figure~\ref{Joos} shows that the first maximum is slightly higher for pyramidal II than I and thus pyramidal-I slip would require less shear stress (2.36 GPa) compared to the pyramidal-II slip (2.65 GPa) in Mg.  

Although the x-Relaxed DFT calculated GSFE curves here can provide evidence of  $\frac{1}{2} \langle c+a \rangle$ as observed in our experiment than those reported earlier from the standard approach and they can provide clues on dislocation core structures, it is still not possible to discern the actual core structure based on these curves alone. For this reason, in what follows, we calculate the configuration and response of the dislocation cores under stress using MD.

\subsection{Full and partial $\langle c+a \rangle$ pyramidal-II dislocations}

As part of the MD model set up, we first checked that the MEAM potential used in the present MD calculations produces the same GSFE curve under x-relaxed boundary conditions. Figure~\ref{slip-py1-py2} compares the MD calculated GSFE curves from both approaches. The standard approach qualitatively gives similar results for MD and DFT (see Figure~\ref{slip-py1-py2}). In the GSFE curves from the x-Relaxed approach, on the other hand, the local minimum for pyramidal-I slip has lower energy compared to the local minima for pyramidal-II slip. This discrepancy indicates that the MEAM potential could be tuned using the x-Relaxed approach to better predict the energetics of these two different slip systems. Notably, however, the x-Relaxed GSFE curve for the MEAM potential provides the local minimum at 0.5 normalized displacement for the pyramidal-II dislocation. 

We also produced GSFE curves for pyramidal-I and pyramidal-II slip with the Sun \cite{Sun-EAM} and Liu \cite{Liu-EAM} embedded atom model (EAM) potentials. Figure~\ref{slip-py1-py2} (c) and (d) compares the GSFE curves from both the standard and x-Relaxed approach.  Similar to DFT and MEAM, the curves from these two other EAM potentials experience a dramatic change when the x-Relaxed approach is applied to permit local shuffling. In essence in both cases, a local minimum lies at 0.5 in the x-Relaxed GSFE curves that was not present with the standard approach.  These results suggest that these MD potentials should predict that full pyramidal type II dislocations could achieve a lower energy state by dissociating into the SP configuration. 

In our high resolution electron microscope analysis, we found a single $\frac{1}{2}\langle c+a \rangle$ partial near a twin boundary on the $(11 {\bar 2} 2)$ slip plane (Figure~\ref{expt-result}). In order to understand the observance of single $\frac{1}{2}\langle c+a \rangle$  pyramidal type II partial, we carried out MD simulations using the MEAM potential for the movement of a pyramidal II dislocation. 

We first simulate the dissociation of a full $\langle c+a \rangle$ edge dislocation on the $(11 {\bar 2} 2)$ slip plane.  The simulation cell consists of 60440 atoms and is periodic in the x direction, which is along the dislocation line, while the surface normals to the y and z directions are free. The cell dimensions are 27.6 $\AA$ $\times$ 245.9 $\AA$ $\times$ 206.7 $\AA$ in the x, y and z directions respectively. We create the full $\langle c+a \rangle$ edge dislocation in the middle of the simulation cell using the elastic Volterra displacement field as shown in Figure~\ref{splitting-edge-py2}(a). Figure~\ref{splitting-edge-py2}(b) shows that in MD, the full dislocation dissociates into two $\frac{1}{2}\langle c+a \rangle$ partials.   

To quantify the separation of the two partials, we perform a disregistry analysis\cite{Pilania, Dholabhai} using two $(11 {\bar 2} 2)$  layers near the dislocation core. To calculate the disregistry, we take two layers of the perfect crystal as the reference and compute disregistry vectors as: 

\begin{equation}
\delta r = \vec{r}_{ij}^{R} -\vec{r}_{ij}^{I}
\end{equation}

\noindent where $\vec{r}_{ij}^{I}$ is the relative position between the $i^{th}$ and $j^{th}$ atoms that form a pair in the reference and $\vec{r}_{ij}^{R}$ is the relative position between the same pair of atoms at the relaxed layers. The disregistry analysis shows that after relaxation, the full $\langle c+a \rangle$ dislocation splits into two partials separated by ~22.6 $\AA$. The dislocation core structures for the partials as well as their separation is similar as reported in the work of Wu et. al. \cite{Curtin1}

Next using MD, we studied the response this SP configuration under stress. The stress is applied such that both the leading and trailing partials have the same resolved shear stress. Figure~\ref{splitting-edge-py2}(c) and (d) shows the configuration after some time under application of strain. Under a suitably applied strain (=0.5$\%$), the leading partial moves first while the trailing remains stationary. The trailing partial starts to move after the leading has moved 3$\AA$. The stacking fault width consequently expands to $\approx 25.6 \AA$ and maintains this distance as they move in concert. This stress-induced response is consistent with the GSFE curve, in which the leading partial has the lower $\gamma_{usf}$ than the trailing partial. When the sense of the applied strain is reversed, the partial dislocations move in the opposite direction, again with the right hand partial moving first (now the trailing partial). The steady-state stacking fault width shrinks from $ \approx 25.6 \AA$ to $\approx 19.6 \AA$. This analysis shows that the SP extended pyramidal II slip dislocation is maintained and is glissile on the $\{11{\bar2}2 \}$ plane in both the forward and reverse glide direction and the separation distance of the two partials changes under positive and negative shear strain. Thus we find asymmetry in the applied strains to move the partial dislocations towards the left and the right, consistent with the asymmetry in the GSFE curve. 

Other mechanical phenomena are also known to split extended dislocation cores. Application of non-glide stresses can cause further widening or narrowing of the stacking fault via the Escaig effect \cite{Wang-Beyerlein}.  Dislocation speeds reaching close to the speed of sound have also been shown to augment the elastic interaction fields between the partials such that the dynamic stacking fault widths deviate from static ones \cite{Weertman, Lothe}. In order to see whether these two partials can be separated further, we performed additional loading conditions, such as non-glide stresses or higher stress levels. However, in neither scenario did the two $ \frac{1}{2}\langle c+a \rangle$ partial dislocations fully split apart. Taken together, we can conclude that the two partials cannot be separated far apart from each other when both are free to move. 

As another explanation for the observed single partial, we studied an isolated $ \frac{1}{2}\langle c+a \rangle$ partial dislocation. We find that whether an isolated $\langle c+a \rangle$ partial is stable depends on the applied stress.  In one MD simulation, we created a single partial dislocation in the middle of simulation cell and found that it moves to the free surface during MD relaxation. This result is not surprising as creation of a single $\langle c+a \rangle$ partial creates a stacking fault and the single partial moves to the free surface to remove the energy penalty incurred by the fault. However, under application of a shear strain, the single partial can be stabilized.

The foregoing analysis would support the following possibility for the origin of the $\frac{1}{2}\langle c+a \rangle$ partial seen experimentally (See Figure~\ref{expt-result}). Through the dissociation reaction of a pyramidal-II dislocation at the ($10{\bar1}2$) twin boundary \cite{Yoo-1,Yoo-2, PRS-Beyerlein, Mendelson-1,Mendelson-2}, the $ \frac{1}{2}\langle c+a \rangle$ partial is produced.  The ($10{\bar1}2$) twin produces an internal strain field \cite{Arul} and under the action of this field, the 1/2 $\langle c+a \rangle$ partial is driven to move a distance into the crystal away from the twin boundary.  Under this local field, the partial is stable. The other product of the reaction lies at the twin boundary and may either be a sessile residual dislocation or the trailing dislocation. To explore this possibility, in another simulation, we pinned the left partial and applied a shear strain on the system parallel to $(11 {\bar 2} 2)$ plane. We find that the right partial can easily move far apart under applied stress as shown in Figure~\ref{explain-expt-result}.

Taken together, the observation of a $\frac{1}{2}\langle c+a \rangle$ partial dislocation on the $\{11{\bar2}2 \}$ plane can be explained by the motion of pyramidal II dislocations, in which one energy minimum state consists of the SP reaction of two $\frac{1}{2}\langle c+a \rangle$ dislocations split in the glide plane. 

\subsection{Comparison to pyramidal-I dislocations}

We also studied the dissociation of mixed and near edge $ \langle c+a \rangle$ dislocations on the pyramidal-I plane using the MEAM potential. Figure~\ref{splitting-py1}(b) shows the result of this MS calculation for the near edge dislocation, which deviates 15.2 $^{\circ}$ from pure $ \langle c+a \rangle$ edge dislocation. The dislocation dissociates on the pyramidal-I plane into two partials with a stacking fault width of 30.4 $\AA$. It can be shown that the mixed full dislocation relaxes to the same core configuration. As discussed earlier, the energy minimum in the GSFE curve for the $\{10{\bar1}1 \}$ plane may not lie on $\{10{\bar1}1 \}$ plane but could be out of the plane. Based on the Burgers circuit drawn around the two dissociated partials, the Burgers vectors of the two partials are $\frac{1}{6} [20{\bar2}3]$ and $\frac{1}{6} [02{\bar2}3]$, which point out of $\{10{\bar1}1 \}$ plane.  These core configuration conforms to experimental observations in an Mg alloy \cite{ Sandlobes}. These Burgers vectors lie on the $(30{\bar3}4)$ and $(30{\bar3}2)$ planes, respectively. As shown in Figure~\ref{splitting-py1}, however, they are located within the $\{10{\bar1}1 \}$ plane. Figure~\ref{gsfe3034} shows the GSFE curve for $(30{\bar3}4)  \langle 20{\bar2}3 \rangle $ slip system obtained from DFT and MEAM. The dissociated configuration of the pyramidal-I dislocation is consistent with this GSFE curve, which displays a local minimum at $\frac{1}{6} [20{\bar2}3]$.

Figure~\ref{splitting-py1} (d) shows the core of the two partials obtained after the dissociation reaction when starting with a full $ \langle c+a \rangle$ dislocation. The core of left partial is spread out on the $\{10{\bar1}1 \}$plane, whereas the core of the right partial is spread on $\{10{\bar1}1\}$, which is consistent with the earlier work.\cite{Curtin1}

We applied a shear stress on the nominal $(10{\bar 1}1)$ plane and in the $[10 {\bar 1} 2]$ direction in order to test the mobility of the extended edge and mixed dislocations. For both dislocations, it is found that right partial is glissile and moves away from the left partial on $(10{\bar 1}1)$ plane, whereas the partial on the left does not move.  This shear in this direction extends the stacking fault between them. Under reverse shear load, however, the partial on the right moves towards the partial on the left and before intersecting it, climbs onto the basal plane. 

\section{Conclusions}

We used first-principles density functional theory to study the generalized stacking fault energy surface for pyramidal-I and pyramidal-II slip systems in Mg. We show that allowing additional relaxation perpendicular to glide results in local shuffling of atoms near the slip planes, which lowers the stable stacking fault energy for pyramidal-II slip compared to the pyramidal-I slip in Mg. The calculated GSFE for pyramidal-II slip with this relaxation approach shows that the Burgers vector for the partials would be exactly $\frac{1}{2}\langle c+a \rangle$. Here using high resolution TEM we also provide evidence of a single $\frac{1}{2}\langle c+a \rangle$ partial dislocation of a pyramidal II dislocation on the the $\{11{\bar2}2 \}$ plane. To explain the observation, we performed MD simulations and found that a full edge $\langle c+a \rangle$ dislocation splits into two partials on the $\{11{\bar2}2 \}$ plane separated by $\approx 22.6 \AA$. Although it is difficult to further separate the two $\frac{1}{2} \langle c+a \rangle $ partials far apart when both are free to move, one partial can move far apart if other is pinned at a twin boundary or other defect. Based on this analysis, we postulate that the HR TEM observation of a single partial near the twin boundary results because one partial is pinned at the twin boundary and the other has moved into the bulk under an internal stress. We also simulated the dissociation of a full edge and mixed $\langle c+a \rangle$ pyramidal I dislocation. These dislocations achieve a lower energy state by dissociating into two partials following $\frac{1}{6} [20{\bar2}3]$ and $\frac{1}{6} [02{\bar2}3]$ lying respectively on $(30{\bar3}4)$ and $(30{\bar3}2)$ planes.  

Simulations of the glide response under an applied shear stress show that an edge pyramidal-II slip dislocation can glide as an extended dislocation on the $(11 {\bar 2} 2)$ plane. The separation distance changes as it moves and depends on the sense of shear. Only one partial of the extended edge and mixed pyramidal I dislocation can move, while the other one remains sessile.    

\section{Acknowledgments}
We would like to thank financial support from Los Alamos National Laboratory Directed Research and Development ER grant 20140348ER and LANL high performance computer facility.


\begin{thebibliography}{}

\bibitem{Lu-1} K. Lu and L. Lu, Scripta Mater {\bf 66}, 835 (2012).
\bibitem{Lu-2} K. Lu, L. Lu, and S. Suresh, Science {\bf 324}, 349 (2009).
\bibitem{Pollock} T. M. Pollock, Science {\bf 328}, 986 (2010).
\bibitem{Cole} G. S. Cole and A. M. Sherman, Materials Characterization {\bf 35}, 3 (1995).
\bibitem{Bauer} J. Bauer, S. Hengsbach, I. Tesari, R. Schwaiger, and O. Kraft, Proceedings of the National Academy of Sciences 111, 2453 (2014).
\bibitem{Williams} J. C. Williams and E. A. Starke Jr, Acta Mater {\bf 51}, 5775 (2003).
\bibitem{Kumar-acta} A. Kumar, J. Wang, and C. N. Tome, Acta Mater {\bf 85}, 144 (2015).
\bibitem{Kumar-apl} A. Kumar, I. J. Beyerlein and J. Wang, App. Phys. Lett. {\bf 105}, 071602 (2014).
\bibitem{HCWu} H. C. Wu, A. Kumar, J. Wang, X. F. Bi, C. N. Tome, Z. Zhang $\&$ S. X. Mao, Scientific Report {\bf 6}, 24370 (2016).
\bibitem{Wonsiewicz} B. C. Wonsiewicz  and W. A. Backofen, W. A., Trans. Metall. Soc. AIME {\bf 239}, 1422 (1967).
\bibitem{RobertCS} C. S. Robert, Magnesium and Its Alloys, John Wiley $\&$  Sons, Inc.: New York, NY, USA (1960).
\bibitem{Obara} T. Obara, H. Yoshinga, S. Morozumi, Acta Metall., {\bf 21}, 845 (1973).
\bibitem{Agnew1} S. R. Agnew, J. A. Horton, M.H. Yoo, Metall. Mater. Trans. A. {\bf 33}, 851 (2002).
\bibitem{Byer} C.M. Byer, B. Li, B. Cao, K. Ramesh, Scr. Mater. {\bf 62}  536 (2010).
\bibitem{Lilleodden} E. Lilleodden, Scr. Mater. {\bf 62}, 532 (2010)
\bibitem{Kelvin} Kelvin Y. Xie, Zafir Alam, Alexander Caffee, Kevin J. Hemker, Scripta Materialia {\bf 112}, 75 (2016).
\bibitem{Sandlobes} S. Sandlobes, M. Friak, J. Neugebauer, D. Raabe, Mater. Sci. Eng., A {\bf 576}, 61 (2013).
\bibitem{Syed} B. Syed, J. Geng, R. Mishra, K. Kumar, Scr. Mater. {\bf 67}, 700 (2012).
\bibitem{Geng} J. Geng, M.F. Chisholm, R. Mishra, K. Kumar,  Philos. Mag. Lett. {\bf  94} , 377 (2014).
\bibitem{Stohr-1972} J. F. Stohr, J. P. Poirier, Phil. Mag. {\bf 25}, 1313 (1972).
\bibitem{Yu-Qian-1} Q. Yu, L. Qi, R. K. Mishra, J. Li, and A. M. Minora,  Proceedings of the National Academy of Sciences {\bf 110} 13289, (2013).
\bibitem{Ando-S} S. Ando, T. Gotoh and H. Tonda, Mettallurgical $\&$ Materials Transactions A {\bf 33}, 823 (2002).
\bibitem{Morris} M. H. Yoo, J. R. Morris, K.M. Ho, and S.R. Agnew, Mettallurgical $\&$ Materials Transactions A {\bf 33}, 813 (2002).
\bibitem{Curtin1} Z Wu, M F Francis and W A Curtin, Modelling Simul. Mater. Sci. Eng. {\bf 23}, 015004 (2015).
\bibitem{CurtinNature} Z Wu, W. A. Curtin, Nature {\bf 526}, 62 (2015)
\bibitem{DH-Kim} D.-H. Kim, F. Ebrahimi,  M.V. Manuel, J.S. Tulenko, S.R. Phillpot, Materials Science and Engineering A {\bf   528}, 5411 (2011).
\bibitem{BLi} B. Li and E. Ma, Philosophical Magazine {\bf  89},  1223 (2009).
\bibitem{Alwady} Y. Tang, J. A. El-Awady, Acta. Matterilia {\bf 71}, 319 (2014).
\bibitem{Guo} Y. F. Guo, X.Z. Tang, Y.S. Wang, Z.D. Wang, S. Yip, Acta Metall. Sin. (Engl. Lett.) {\bf 26}, 75 (2013).
\bibitem{Curtin2} M. Ghazisaeidi, L.G. Hector Jr, W.A. Curtin, Scripta Materialia {\bf 75},  42, (2014).
\bibitem{Nogaret} T. Nogaret, W. Curtin, J. Yasi, L. Hector, D. Trinkle, Acta Mater. {\bf 58}, 4332 (2010).
\bibitem{perdew} J. P. Perdew and A. Zunger, Phys. Rev. B {\bf 23}, 5048 (1981).
\bibitem{vasp1} G. Kresse and J. Hafner, Phys. Rev. B {\bf 49}, 14251 (1994). 
\bibitem{vasp2} G. Kresse and J. Furthm\"{u}ller, Phys. Rev. B {\bf 54}, 11169 (1996). 
\bibitem{paw1} P. E. Blöchl, Phys. Rev. B {\bf 50}, 17953 (1994).
\bibitem{paw2}  G. Kresse and D. Joubert, Phys. Rev. B {\bf 59}, 1758 (1999).
\bibitem{mh-pack} J. D. Pack and H. J. Monkhorst, Phys. Rev. B {\bf 13}, 5188 (1976); ibid Phys. Rev. B {\bf 16}, 1748 (1977).
\bibitem{lammps}  S. Plimpton, J. Comput. Phys. {\bf 117}, 1 (1995).
\bibitem{Chaari} N. Chaari, E. Clouet , D. Rodney, Metallurgical and Materials Transactions A {\bf 45}, 5898 (2014).
\bibitem{Bacon} D.J. Bacon, M.H. Liang, Philos. Mag. A {\bf 53}, 163 (1986).
\bibitem{Jones} I.P. Jones, W.B. Hutchinson, Acta Metall. {\bf 29}, 951 (1981).
\bibitem{Q-Zu} Q. Zu et al.: Acta Metall. Sin. (Engl. Lett.), {\bf 28} , 876 (2015).
\bibitem{Joos-Duesbery} B. Joos, M. Duesbery , Phys. Rev. Lett. {\bf 78}, 266 (1997)
\bibitem{Sun-EAM} D.Y. Sun, M.I. Mendelev, C.A. Becker, K. Kudin, T. Haxhimali, M. Asta, J.J. Hoyt, A. Karma, and D.J. Srolovitz, Phys. Rev. B {\bf 73}, 024116 (2006).
\bibitem{Liu-EAM} X-Y. Liu, J. B. Adams, F. Ercolessi and J. A Moriarty, Modelling Simul. Mater. Sci. Eng. {\bf 4},  293 (1996).
\bibitem{Pilania} G. Pilania, B. J. Thijsse, R. G. Hoagland, I. Lazic, S. M. Valone, and X.-Y. Liu, Sci. Rep. {\bf 4}, 4485 (2014).
\bibitem{Dholabhai}  P. Dholabhai,	G. Pilania,	J. A. Aguiar,	A. Misra	$\&$ B. P. Uberuaga, Nature Communications {\bf 5}, 5043 (2014). 
\bibitem{Wang-Beyerlein} Z. Q. Wang and I. J. Beyerlein, Phys. Rev. B {\bf 77}, 184112 (2008)
\bibitem{Weertman} J. Weertman, in Response of Metals to High Velocity Deformation, edited by P. G. Shewmon and V. F. Zackay (Interscience, New York, 1961).
\bibitem{Lothe} J. P. Hirth and J. Lothe, in Dislocation Dynamics, edited by A. R. Rosenfield, G. T. Hahn, A. L. Bement, Jr., and R. I. Jaffee (McGraw-Hill, New York), 231 (1967)
\bibitem{Yoo-1} M. H. Yoo, Trans. Metall. Soc. AIME {\bf 245}, 2051 (1969).
\bibitem{Yoo-2} M. H. Yoo, Metall. Mater. Trans. A {\bf 12}, 409  (1981).
\bibitem{PRS-Beyerlein} I. J. Beyerlein, J. Wang, M. R. Barnett, C. N. Tomé, Proc. R. Soc. A {\bf 468}, 1496 (2012).
\bibitem{Mendelson-1} S. Mendelson,  Mater. Sci. Eng. {\bf 4}, 231 (1969).
\bibitem{Mendelson-2} S. Mendelson, J. Appl. Phys. {\bf 41}, 1893  (1970).
\bibitem{Arul} M. Arul Kumar, A.K. Kanjarla, S.R. Niezgoda, R.A. Lebensohn and C.N. Tome, Acta Materialia {\bf 84}, 349 (2015).
\bibitem{Pradhan} G. K. Pradhan, A. Kumar, S. K. Deb, U. V. Waghmare, C. Narayana, Physical Review B {\bf 82}, 144112 (2010)

\end{thebibliography}
\end{document}